\documentclass[sigconf]{acmart}
\copyrightyear{2026}
\acmYear{2026}
\setcopyright{cc}
\setcctype{by}
\acmConference[SIGIR '26] {Proceedings of the 49th International ACM SIGIR Conference on Research and Development in Information Retrieval}{July 20--24, 2026}{Melbourne, VIC, Australia.}
\acmBooktitle{Proceedings of the 49th International ACM SIGIR Conference on Research and Development in Information Retrieval (SIGIR '26), July 20--24, 2026, Melbourne, VIC, Australia}
\acmISBN{979-8-4007-2599-9/2026/07}
\acmDOI{10.1145/3805712.3808443}

\usepackage{graphicx} 
\usepackage{xcolor}
\usepackage{listings}
\usepackage{caption}
\usepackage{hyperref}
\usepackage{orcidlink}
\usepackage{enumitem}

\lstdefinestyle{custom}{
  basicstyle=\ttfamily\small,
  backgroundcolor=\color{white},
  keywordstyle=\color{blue},
  commentstyle=\color{gray},
  showstringspaces=false,
  breaklines=true,
  moredelim=**[is][\colorbox{yellow}]{@}{@}
}

\keywords{Dynamic Facet Suggestions, Job Search, Small Language Models}

\title{Policy-Grounded Dynamic Facet Suggestions for Job Search}

\author{Dan Xu}
\authornote{Equal contribution.}
\affiliation{
 \institution{LinkedIn Corporation}
 \city{Mountain View}
 \state{CA}
 \country{USA}
}
\email{dnxu@linkedin.com}

\author{Baofen Zheng}
\authornotemark[1]
\affiliation{%
  \institution{LinkedIn Corporation}
  \city{Mountain View}
  \state{CA}
  \country{USA}}
\email{bzheng@linkedin.com}

\author{Qianqi Shen}
\affiliation{%
  \institution{LinkedIn Corporation}
  \city{Mountain View}
  \state{CA}
  \country{USA}}
\email{qishen@linkedin.com}

\author{Jianqiang Shen}
\affiliation{%
  \institution{LinkedIn Corporation}
  \city{Mountain View}
  \state{CA}
  \country{USA}}
\email{jshen@linkedin.com}

\author{Wenqiong Liu}
\affiliation{%
  \institution{LinkedIn Corporation}
  \city{Mountain View}
  \state{CA}
  \country{USA}}
\email{ecliu@linkedin.com}

\author{Chunnan Yao}
\affiliation{%
  \institution{LinkedIn Corporation}
  \city{Mountain View}
  \state{CA}
  \country{USA}}
\email{chyao@linkedin.com}

\author{Ping Liu}
\affiliation{%
  \institution{LinkedIn Corporation}
  \city{Mountain View}
  \state{CA}
  \country{USA}}
\email{piliu@linkedin.com}

\author{Rajat Arora}
\affiliation{%
  \institution{LinkedIn Corporation}
  \city{Mountain View}
  \state{CA}
  \country{USA}}
\email{rajarora@linkedin.com}

\author{Kevin Kao}
\affiliation{%
  \institution{LinkedIn Corporation}
  \city{Mountain View}
  \state{CA}
  \country{USA}}
\email{kkao@linkedin.com}

\author{Hsiang Lin}
\affiliation{%
  \institution{LinkedIn Corporation}
  \city{Mountain View}
  \state{CA}
  \country{USA}}
\email{hslin@linkedin.com}

\author{Wanjun Jiang}
\affiliation{%
  \institution{LinkedIn Corporation}
  \city{Mountain View}
  \state{CA}
  \country{USA}}
\email{wanjiang@linkedin.com}

\author{Yusuke Takebuchi}
\affiliation{%
  \institution{LinkedIn Corporation}
  \city{Mountain View}
  \state{CA}
  \country{USA}}
\email{ytakebuchi@linkedin.com}

\author{Jingwei Wu}
\affiliation{%
  \institution{LinkedIn Corporation}
  \city{Mountain View}
  \state{CA}
  \country{USA}}
\email{jingwu@linkedin.com}

\author{Wenjing Zhang}
\affiliation{%
  \institution{LinkedIn Corporation}
  \city{Mountain View}
  \state{CA}
  \country{USA}}
\email{wzhang@linkedin.com}

\begin{document}

\begin{abstract}
Job seekers often initiate search with short, underspecified queries. At LinkedIn, over $80\%$ of job-related queries contain three or fewer keywords, making accurate user intent inference and relevant job retrieval particularly challenging.
We present \emph{dynamic facet suggestion} (DFS), an interactive query-refinement mechanism that facilitates intent disambiguation by surfacing personalized semantic attributes conditioned on the joint user-query context in real time. We propose a policy‑grounded, retrieval‑augmented ranking framework for facet suggestion, comprising offline taxonomy curation, embedding‑based retrieval of top‑$K$ candidates,  and a  distilled small language model (SLM) based candidate scoring. The system is optimized for real-time serving via  point-wise single-token scoring and batching/prefix caching. Offline evaluation demonstrates high precision for generated suggestions, and online A/B tests show significant lifts in suggestion engagement and  job search outcomes. 
\end{abstract}

\ccsdesc[500]{Information systems~Retrieval models and ranking}
\ccsdesc[500]{Information systems~Users and interactive retrieval}

\maketitle

\begin{figure*}[tb]
    \centering
    \includegraphics[width=0.83\textwidth]{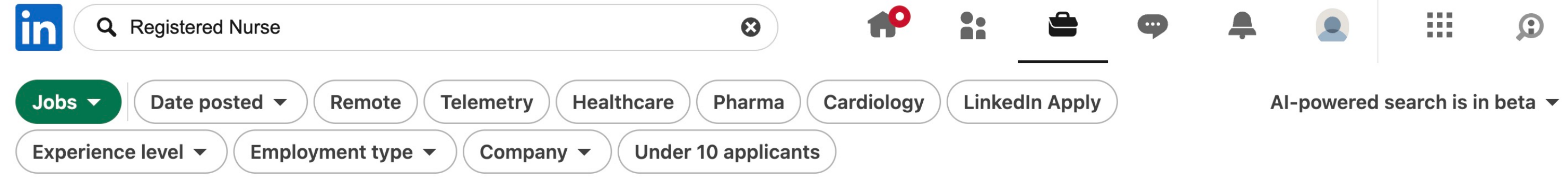}
    \vspace{-4pt}
    \caption{Example of dynamic facet suggestion in AI-powered job search for the query \emph{Registered Nurse}. The system dynamically suggests \emph{Remote} as the workplace type, \emph{Healthcare} and \emph{Pharma} as industry facets, and \emph{Telemetry} and \emph{Cardiology} as domain knowledge facets. Other facets are statically defined and not model-driven.}
    \vspace{-3pt}
    \label{fig.example}
       \Description{Dynamic facet suggestion for the query Registered Nurse, including workplace type, industry, and domain knowledge facets.}
\end{figure*}

\section{Introduction}

Job search is a high-stakes, intent-driven retrieval problem where user queries are often short and ambiguous. In practice, many job seekers begin with underspecified queries  and must iteratively refine them to converge toward more actionable results. In LinkedIn's AI-powered job search system, over 80\% of queries contain three or fewer keywords, and a substantial fraction of user feedback indicates that members require additional guidance to effectively articulate their intent.
We introduce \emph{dynamic facet suggestion} (DFS),  which surfaces a  set of contextualized, query-aware attributes (e.g., workplace type, industry, function, domain knowledge) directly beneath the search bar, as illustrated in Figure~\ref{fig.example}. Unlike static dropdown filters such as \emph{Experience level} and \emph{Employment type}, these facet ``pills'' are dynamically generated and personalized based on joint member–query context. When selected, a suggested facet is appended to the original query, transforming an underspecified query into a more expressive next query and enabling incremental construction of multi-attribute intent with minimal user effort.

Conceptually, DFS can be viewed as a form of \emph{interactive query reformulation}: the system proposes a small set of candidate refinements and the user selects one to form the next query. This formulation is related to query expansion and rewriting, where classical techniques include relevance feedback and pseudo-relevance feedback (PRF) \cite{rocchio1971relevance,lavrenko2001relevance}, while recent work uses large language models (LLMs) to generate expansions or rewrites \cite{qe_llm_Rolf,qe_llm_query_to_doc}. However, these approaches are not directly suitable for our setting. Unconstrained generation can produce hallucinated or policy-violating suggestions, while PRF can amplify retrieval errors when the initial query is ambiguous. Moreover, user-facing job search requires high precision and strict latency guarantees \cite{kenthapadi2017personalized, LinkRetrieval2024}. 

Driven by these constraints, we propose a \emph{policy‑grounded} retrie- val‑and‑ranking framework for DFS that operates over a curated facet taxonomy. Deploying the DFS system at production scale poses two key challenges. First, suggested facets must satisfy product policy quality criteria that jointly consider query relevance, member relevance, and refinement utility. Second, the system must scale to high concurrency under strict latency and cost constraints, necessitating efficient candidate generation and ranking. We address these challenges and make the following contributions:
\begin{itemize}[leftmargin=*, itemsep=0pt, topsep=2pt, label={}]
\item \textbf{Policy-Driven Evaluation Framework:} We introduce a relevance formulation and annotation schema that evaluates suggestions based on product policy. It integrates query relevance, member alignment, and refinement utility into a unified metric. 
\item \textbf{Latency-Optimized Architecture:} We propose a scalable system that integrates taxonomy curation, an embedding SLM-based candidate retriever (EBR), and a text generation SLM-based ranking module with batching/prefix caching for low-latency serving. 
\item \textbf{Advanced SLM Fine-tuning:} We present methodologies for building high‑performing SLMs, including a LLM‑as‑a‑judge generated high quality training data, product‑policy–aligned prompting, and a multi‑stage fine‑tuning pipeline.
\item \textbf{Production-Scale Evaluation:} We provide offline evaluation metrics and online A/B experimentation results, demonstrating significant improvements in user engagement and search efficiency. Additionally, we discuss the impact of query-distribution shifts induced by interactive query reformulation and their implications and potential mitigations.
\end{itemize}


\section{Related Work}
\label{sec.related_work}

\emph{Query expansion and query rewriting.}
Classical query expansion includes relevance feedback \cite{rocchio1971relevance} and pseudo-relevance feedback using relevance models \cite{lavrenko2001relevance}, as well as lexical and semantic expansion \cite{qe_sigir_93,qe_sigir_94}. Neural network based approaches leverage distributed representations and contextual encoders to generate or select expansion terms \cite{word_embeddings_16,word_embeddings_19,zheng-etal-2020-bert}.

\emph{LLM-based query expansion.}
Recent work explores prompting or fine-tuning LLMs for query expansion and generative relevance feedback \cite{qe_llm_Rolf,qe_llm_Iain,qe_llm_query_to_doc,qe_llm_Corpus,qe_llm_GaQR}, while also documenting failure modes such as hallucination and brittleness on ambiguous queries \cite{qe_llm_Abe}. Our work differs by focusing on an interactive query$\rightarrow$next-query refinement interface and by constraining suggestions to a curated facet taxonomy to improve controllability.

\emph{Faceted search and interactive refinement.}
Faceted search interfaces support iterative narrowing through structured attributes \cite{yee2003faceted,hearst2006design,tunkelang2009faceted}. DFS can be viewed as a lightweight in-context faceted refinement mechanism that adapts the suggested facets to the current query and member context,  rather than exposing a uniform set of static dropdown filters to all users.

\emph{Policy-guided generation and production LLM systems.}
Alignment work has explored using explicit principles or policies (``constitutions'') to guide model outputs \cite{bai2022constitutional}. In our setting, a product policy plays the role of a domain-specific constitution that defines acceptable refinements. From a systems perspective, our architecture follows retrieval-augmented generation patterns \cite{lewis2020retrieval} and relies on efficient LLM serving with batching and prefix caching \cite{kwon2023vllm}. We complement production LLM systems for job search query understanding \cite{liu2025powering} by focusing on interactive refinement and by analyzing query-distribution shift induced by the refinement UI.

\section{Policy-grounded Facet Suggestion Framework}
\label{sec.framework}

In this section, we introduce the policy-grounded relevance framework to facilitate DFS within the LinkedIn job search ecosystem.

\subsection{Problem Definition}
Given a search query \(q\) and member context \(m\) (e.g., preferred job titles and industry), our goal is to produce a ranked list of $K$ facet suggestions \(\{(t_i, v_i)\}_{i=1}^{K}\), where \(t_i\) is the facet type (e.g., workplace type, industry) and \(v_i\) is the suggested value. In the production interface, selecting a facet suggestion appends the corresponding attribute to the original query, inducing a state transition from \( q \) to a refined query \( q' \). We focus exclusively on \emph{narrowing refinements} which increase query specificity without altering its core intent. Formally, we require refinements to satisfy a \emph{monotonic precision} property: the intent expressed by \( q' \) must constitute a strict subset of the intent expressed by \( q \).

While alternative interaction paradigms may permit broadening refinements (e.g., disjunctive expansions) or full query rewrites, such operations frequently introduce semantic drift and violate users’ expected refinement utility. We therefore constrain the action space to refinements that preserve the user’s original intent and leave richer rewriting-based refinements to future work.

\begin{figure*}[tb]
    \centering
    \includegraphics[width=0.83\textwidth]{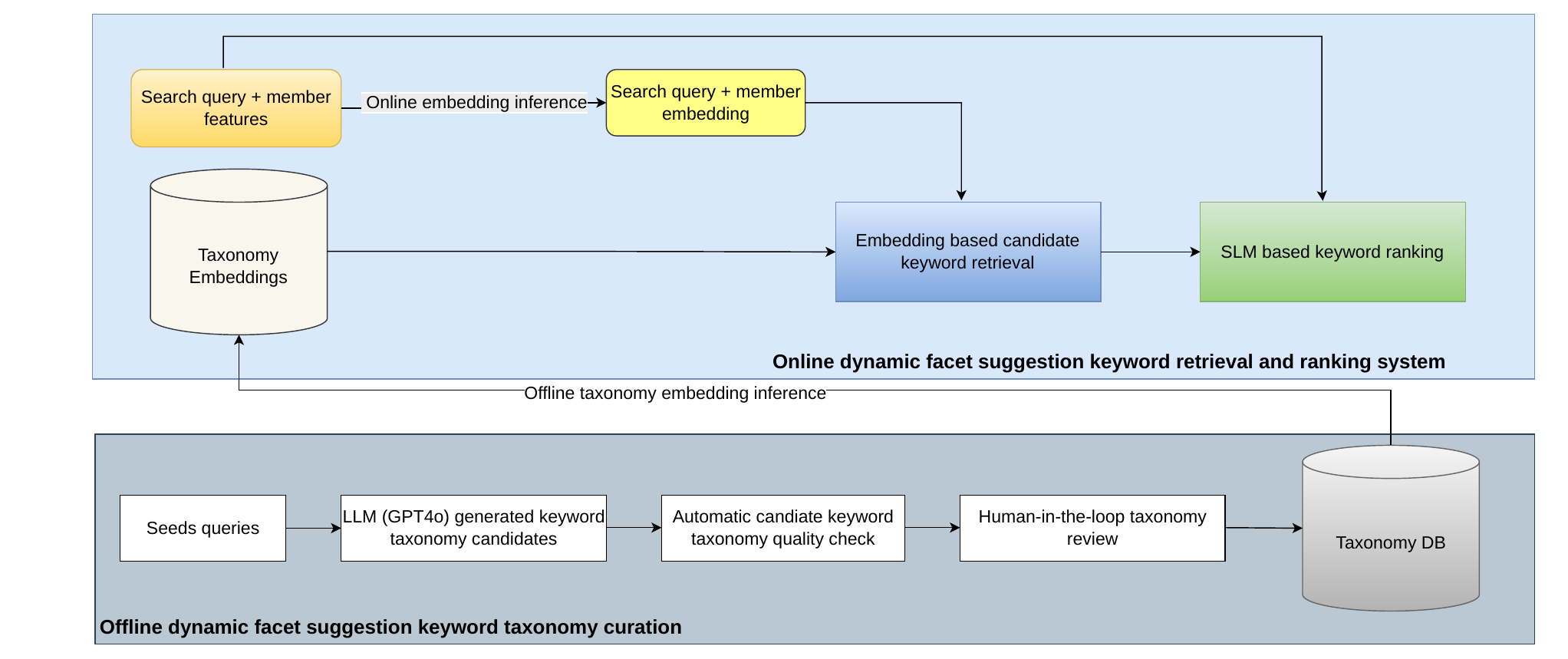}
    \vspace{-4pt}
    \caption{End-to-end dynamic facet suggestion workflow, including the offline LLM-powered taxonomy curation pipeline and the online facet keyword retrieval and ranking system.}
    \vspace{-3pt}
    \label{fig.flow}
       \Description{End-to-end dynamic facet suggestion workflow with offline taxonomy curation and online keyword retrieval and ranking.}
\end{figure*}

\subsection{Product Policy and LLM-as-a-Judge}
\label{sec-llm-judge}

Standard relevance metrics are insufficient for capturing the nuanced product and business constraints inherent in professional search domains such as job discovery. To address this gap, we define a Policy-Grounded Relevance framework that evaluates each candidate suggestion against three core axioms, shown in Table~\ref{tab-product-policy}. We operationalize these axioms using a binary relevance label (\textit{Okay} vs.\ \textit{Poor}). Liquidity constraints (ensuring sufficient job results after refinement) are enforced separately through downstream gating.
To overcome the cost and scalability limitations of human annotation, we adopt an \textit{LLM-as-a-judge} paradigm \cite{son2024llm, li2025generation} for synthetic labeling and evaluation. We encode the product policy into a high-dimensional prompt space through an iterative workflow:

\begin{itemize}[leftmargin=*, itemsep=0pt, topsep=2pt, label={}]
\item \textbf{Expert Benchmarking}: Domain experts (Product Managers) curate a gold-standard seed dataset with intra-pair stability \cite{feng2025we}.
\item \textbf{Prompt Optimization}: We iteratively refine the judging prompt by incorporating few-shot exemplars and explicit edge-case definitions, aligning GPT-4o’s latent reasoning with the three relevance axioms.
\item \textbf{Distillation and Scaling}: Once alignment with human judgment reaches acceptable reliability (measured via inter-annotator agreement metrics such as Cohen’s Kappa) \cite{feng2025we}, we deploy the model to generate labels at scale. These labels serve both as a high-precision evaluation benchmark and as a source of distant supervision for training lightweight production models.
\end{itemize}

\begin{table}[t]
\centering
\caption{Facet Suggestion Constitution (rubrics). A suggestion is labeled \textit{Okay} if and only if it satisfies C1--C3.}
\vspace{-6pt}
\begin{tabular}{p{0.10\linewidth}@{\hspace{2pt}}p{0.82\linewidth}}
\toprule
\textbf{C1} & \textbf{Query faithfulness:} the suggested facet value is semantically compatible with the query intent (no contradiction; not a different occupation). \\
\textbf{C2} & \textbf{Member plausibility:} when member context is available and is relevant to the search query, the suggestion is consistent with the member profile or represents a plausible career transition. \\
\textbf{C3} & \textbf{Refinement utility:} the suggestion meaningfully narrows results,  avoids redundancy with the query, and is actionable via append-to-query interaction. \\
\bottomrule
\end{tabular}
\vspace{-6pt}
\label{tab-product-policy}
\end{table}

\subsection{Human-in-the-Loop Taxonomy Curation}

Facet keywords are terms that capture core job attributes, such as job function, workplace type, industry, and domain‑specific specialties essential to a given role. To build our keyword taxonomy, we start from a set of seed queries consisting of (1) parent occupations, which represent the highest level job title taxonomy defined by LinkedIn, and (2) short job search queries containing no more than 3 words. For each seed, we prompt GPT-4o to induce a candidate set of facet keywords.
We then apply a three-dimensional quality control process to evaluate the generated candidates:
\begin{itemize}[leftmargin=*, itemsep=0pt, topsep=2pt, label={}]
\item \textbf{Product policy compliance:} We utilize the LLM-as-a-judge framework (detailed in Section~\ref{sec-llm-judge}) to assess the semantic validity of each keyword and produce an \textit{Okay}/\textit{Poor} judgment.
\item \textbf{Job liquidity validation:} We append each keyword to its corresponding seed query and issue a call to the job search stack. We then verify the returned job count to ensure that the keyword yields a sufficient volume of job postings.
\item \textbf{Search experience alignment:} To avoid overly niche keywords and ensure alignment with common user behavior, we measure the popularity of each expanded query (i.e., seed query + keyword) based on LinkedIn job search traffic. Candidate keywords whose expanded queries exhibit low popularity are filtered out.
\end{itemize}

While filtering low-popularity expanded queries ensures job liquidity and a consistent search experience for the majority of users, it may reduce coverage for niche job intents in the long tail. Future work will investigate taxonomy curation and adaptive filtering strategies to better balance facet consistency with personalized support for specialized queries.

Following the automated validation filters described above, candidate keywords undergo a Human-in-the-Loop review by domain experts to ensure production-grade ontological quality. Validated entries are then ingested into the taxonomy pool,  where we utilize GPT-4o for entity resolution and alias mapping to resolve semantic redundancies. This semi-automated approach effectively balances the generative capacity of LLMs with expert oversight, enabling continuous improvements in facet coverage while maintaining the high precision required for professional search environments. 

We illustrate the end-to-end dynamic facet suggestion workflow in Figure~\ref{fig.flow}, which comprises the offline LLM-powered taxonomy curation pipeline and the online dynamic facet keyword retrieval and ranking system, described next.

\subsection{Embedding SLM based Facet Candidate Retrieval}

Following the standard two-stage retrieval-and-ranking paradigm \cite{lewis2020retrieval}, we fine-tuned  the gte-Qwen2-1.5B-instruct \cite{Li2023TowardsGTE} model   to perform Embedding-Based Retrieval (EBR) \cite{huang2020embedding, juan2025scaling}. We adopt a Siamese  architecture where the same model encodes both query-side features—comprising the search query string and member context attributes—and facet keyword-side representations. To enhance embedding discriminability, we utilize semantic augmentation, incorporating both the keyword name and its natural language definition into the encoder. Training labeling is derived from our LLM-as-a-judge system, ensuring consistent policy alignment and high-fidelity supervision across the entire pipeline.

Empirical analysis reveals that the selection of the optimization objective is critical for aligning the latent representations of queries and facets. Our results show that InfoNCE Loss \cite{oord2018representation} (via in-batch negative sampling) significantly outperforms the baseline. In contrast, alternative objectives (including Binary Cross-Entropy, Pairwise Triplet, ListNet \cite{cao2007learning}) yielded inferior results, likely due to their limited capacity to contrast against the dense negative signal provided by the local batch. The InfoNCE objective is defined as:
\[
\mathcal{L}_{\text{InfoNCE}} = - \frac{1}{N} \sum_{i=1}^{N} \log \frac{\exp\left( \frac{\mathbf{q}_i^\top \mathbf{d}_i}{\tau} \right)}{ \sum_{j=1}^{N}\exp\left( \frac{\mathbf{q}_i^\top \mathbf{d}_j}{\tau} \right)},
\label{infonce} 
\]
where $N$ is the batch size, $\mathbf{q}_i$ and $\mathbf{d}_i$ represent the query and positive facet embeddings for the $i$-th sample, $j$ indexes every facet in a batch, and  $\tau$ is the temperature parameter.

\subsection{SLM based Keyword Ranking}

We formulate the facet ranking task as a generative pointwise classification problem. Given a query \(q\), member features (similar to those used in the EBR model), and a  candidate facet keyword \(c\), a text generation SLM generates a binary response (``Yes'' or ``No''), indicating whether \(c\) is an appropriate keyword to expand \(q\).  To enable keyword ranking, we leverage the probability of the “Yes” token generated by the SLM, denoted as  $p_{\text{Yes}}$ , as the ranking score.

An alternative approach is to fine-tune an SLM to directly generate a ranked list of suggested facet keywords. However, as shown later, this approach incurs substantial auto-regressive decoding overhead, resulting in prohibitive inference latency compared to our method, where only a single token is generated per sample.

We chose the Qwen2.5-1.5B-Instruct \cite{bai2023qwen} as our backbone SLM  as a trade-off between serving cost and performance, GPT-4o and Qwen3-14B \cite{yang2025qwen3} as teacher models, and fine-tuned the SLM through a progressive multi-stage pipeline.

\textbf{Stage 0}: Similar to EBR model training, 
we leverage our LLM-as-a-judge system to generate a high-precision training corpus. This dataset includes both binary labels and Chain-of-Thought (CoT) \cite{wei2022chain} rationales for candidates.

\textbf{Stage 1}: We perform supervised fine-tuning on both Qwen2.5-1.5B-Instruct and Qwen3-14B. By incorporating CoT rationales, we encourage the models to learn the underlying policy logic rather than simple label correlations.

\textbf{Stage 2}: 
To bridge the performance gap between the parameter-constrained student and the teacher, we conduct on-policy distillation \cite{hinton2015distilling,hsieh2023distilling}. The student generates trajectories (predictions and rationales) which are graded by the 14B teacher. This approach mitigates exposure bias by forcing the student to learn from its own error distribution while approximating the teacher’s latent reasoning quality.
Formally, the student learns from the teacher’s feedback on its own generations by minimizing the following on-policy distillation loss:
\[
\mathcal{L}
=
\mathbb{E}_{(x, y_{1:T}) \sim q_\phi}
\left[
\sum_{t=1}^{T}
D_{\text{KL}}\!\left(
p_\theta(y_t \mid x, y_{<t})
\;\|\;
q_\phi(y_t \mid x, y_{<t})
\right)
\right]
\]
\noindent \textbf{where} $x$ is the input prompt, and
\begin{itemize}[leftmargin=*, itemsep=0pt, topsep=2pt, label={}]
    \item $y_{1:T} = (y_1, \dots, y_T)$: response sequence of length $T$, sampled from the student’s own distribution (\textit{on-policy} sampling).
    \item $p_\theta(y_t \mid x, y_{<t})$: teacher model’s conditional token distribution (parameterized by $\theta$).
    \item $q_\phi(y_t \mid x, y_{<t})$: student model’s conditional token distribution (parameterized by $\phi$).
    \item $D_{\text{KL}}(p \| q) = \sum_i p(i) \log \frac{p(i)}{q(i)}$: forward Kullback–Leibler (KL) divergence measuring the difference between teacher and student token distributions.
    \item $\mathbb{E}_{(x, y_{1:T}) \sim q_\phi}[\cdot]$: expectation of KL-divergence loss over tokens of responses $y$. 
\end{itemize}

\textbf{Stage 3}: We removed the reasoning traces from the training labels, retaining only the final class labels, and compressed the input prompt to one-third of its original length. The Stage~2 student model was then further fine-tuned in this compressed space, allowing it to retain the performance while operating with minimal overhead.

\subsection{Model Serving}
To maximize inference throughput, we implement client-side request batching, grouping multiple facet keyword candidates for a single query into a unified execution block. By exploiting the invariant prompt prefix (comprising instructions, member context, and the search query), it allows vLLM \cite{kwon2023vllm} to fully leverage prefix caching. 
With this serving stack, our point‑wise ranking model achieves significantly lower latency than a list‑wise approach. Although list‑wise ranking uses a single prompt, it requires generating many keyword tokens, making it less efficient from a token and latency standpoint.

\section{Evaluation}
\label{sec.evaluation}



\subsection{Offline Model Evaluation}

We conducted a series of experiments to evaluate the effectiveness of the proposed training pipeline, using the LLM-as-a-Judge framework for consistent evaluation.

\textbf{Overall Performance.}
The EBR model retrieves 27 candidates—comprising 16 domain knowledge, 5 functional, 5 industry, and 1 workplace type keywords—from a pool of approximately 400 candidates, based on product policy enforcing facet‑type diversity and offline precision analysis, which are subsequently ranked by the SLM.
In production, we serve the top‑5 ranked facets per query that receive a “Yes” output token from the SLM. The end-to-end models achieve an F1 score of 0.84, and  91.1\% precision for the top-5 served facets. Incremental gains from each training strategy are summarized in Table~\ref{tab:training-performance}. We next analyze the contribution of individual components to the SLM ranking model.

\textbf{Training Stage Performance.}
Starting from the Qwen2.5-1.5B baseline, we first apply supervised fine-tuning with reasoning traces, which is critical for learning policy-aligned semantic relevance and improves the F1 score from 0.48 to 0.76. We then distill knowledge from a fine-tuned Qwen3-14B teacher model into the Qwen2.5-1.5B student, yielding a substantial further gain in performance (F1 = 0.85). Finally, we fine-tune the model using class labels only with compressed prompts, enabling single-token outputs and low-latency inference while largely preserving accuracy (F1 = 0.84) and achieving favorable precision–throughput trade-offs.

\begin{table}[t]
\centering
\caption{Performance evolution across training stages.}
\label{tab:training-performance}
\vspace{-6pt}
\begin{tabular}{lc}
\toprule
\textbf{Training Stage} & \textbf{F1 Score} \\
\midrule
SLM Base (1.5B) & 0.48 \\
+ Reasoning SFT & 0.76 \\
+ Distillation (14B $\rightarrow$ 1.5B) & 0.85 \\
+ Prompt Compressed Class-only SFT & 0.84 \\
\bottomrule
\end{tabular}
\vspace{-6pt}
\end{table}

\textbf{List-wise vs.\ Point-wise based ranking.}
We compare list-wise and point-wise based ranking in terms of relevance quality and latency. The point-wise approach  scores all 27 candidates using single-token generation with separate prompts, whereas the list-wise model jointly evaluates all candidates within a single prompt via multi-token generation. While both formulations achieve comparable relevance quality ($\sim$$90\%$ precision for the top-5 facets), the list-wise approach incurs higher latency that scales with output length as shown in Table~\ref{tab:listwise-pointwise-latency}. In contrast, the point-wise formulation enables efficient batching and predictable latency, achieving a $3\times$$\sim$$4\times$ reduction in P95 latency without increasing GPU cost.

This efficiency arises from vLLM prefix caching: shared instruction, search query, and member features across facet keyword-specific prompts enable reuse of KV states computed during the prefill stage across batched requests, avoiding redundant attention over the common prompt body. Combined with single-token decoding, this amortizes prefill overhead and reduces latency; we therefore adopt the point-wise formulation for subsequent experiments and production deployment.

\begin{table}[t]
\centering
\caption{End-to-end latencies for different formulations.}
\label{tab:listwise-pointwise-latency}
\vspace{-6pt}
\begin{tabular}{lcc}
\toprule
\textbf{Formulation} & \textbf{P95 Latency} & \textbf{Remarks} \\
\midrule
List-wise & 451ms & Multi-token generation \\
Point-wise & 150ms & Single-token scoring \\
\bottomrule
\end{tabular}
\vspace{-6pt}
\end{table}

\begin{table}[t]
\centering
\caption{Latency for real-time facet suggestion.}
\label{tab:serving}
\vspace{-6pt}
\begin{tabular}{lccc}
\toprule
\textbf{Component} & \textbf{QPS per GPU}  &  \textbf{P95 Latency} \\
\midrule
EBR  & $\sim40$  & $\sim$40ms \\
SLM scoring & $\sim15$ & $\sim$150ms \\
\bottomrule
\end{tabular}

\vspace{4pt}
\footnotesize
We use NVIDIA A100 GPU for  serving. For SLM scoring, one request has 27 candidates.
\vspace{-2pt}
\end{table}

\subsection{Online A/B testing results}

Through prompt compression and client-side batching with prefix caching, we reduce prompt tokens by approximately ($3\times$) and amortize shared-prefix prefill across 27 candidates, achieving $\sim$150ms P95 scoring latency under vLLM (Table~\ref{tab:serving}). In online A/B testing, the proposed system improves both refinement engagement and downstream job search outcomes (Table~\ref{tab:online-metrics}). DFS significantly increases facet interaction (e.g., Facet CTR and action count) and yields consistent gains in downstream efficiency signals, including apply-to-viewport ratio and successful job search sessions.

\begin{table}[t]
\centering
\caption{Online A/B metrics for DFS.}
\label{tab:online-metrics}
\vspace{-6pt}
\begin{tabular}{lcc}
\toprule
\textbf{Metric} & \textbf{Lift (\%)}  & \textbf{p-value} \\
\midrule
Overall Facet CTR & +34.8 & $<0.0001$ \\
Job Apply-to-Impression Ratio & +2.6 & 0.0031 \\
Successful Job Search Sessions & +1.6 & <0.0001 \\
\bottomrule
\end{tabular}

\vspace{4pt}
\footnotesize
\textbf{Metric definitions.}
CTR measures the click-through rate of suggested facets. The baseline of overall facet refers to the existing non-dynamic facets.
Apply-to-Impression Ratio measures the likelihood of an apply given an impression.
Successful Sessions denote sessions with $\geq1$ high-intent action (e.g., apply / save).
\vspace{-6pt}
\end{table}

DFS also alters how members express intent by increasing the prevalence of multi-attribute queries (e.g., title + industry + workplace type). While this shift improves refinement engagement, it may introduce out-of-distribution query patterns for downstream models, occasionally leading to transient relevance regressions until those models adapt. This observation highlights the importance of co-training downstream search components on the evolving query distribution induced by interactive refinement systems.

\section{Conclusion and Future Work}
\label{sec.conclusion}

In this paper, we presented a policy-grounded framework for DFS in LinkedIn Job Search. The system combines LLM-assisted construction of facet taxonomies with a fine-tuned embedding SLM for efficient candidate retrieval, followed by a text generation SLM for policy-aligned facet ranking. Our point-wise single-token based scoring design, together with batching and prefix caching, enables production-grade latency at scale. Online A/B experiments show that DFS significantly increases refinement engagement and improves downstream job search outcomes. We also observe that query refinement can shift the query distribution toward multi-attribute patterns, motivating future work on co-training downstream retrieval and ranking models to maintain relevance under evolving query behavior. In addition, we will expand the taxonomy and investigate adaptive candidate filtering strategies to better balance facet consistency with personalized support for specialized queries. Overall, this work provides a practical blueprint for deploying controlled, policy-aligned SLM components in high-throughput, latency-constrained search environments.
\clearpage
\balance

\bibliographystyle{ACM-Reference-Format}
\bibliography{main}

@inproceedings{qe_sigir_93,
  title={Concept based query expansion},
  author={Yonggang Qiu and Hans-Peter Frei},
  booktitle={Proceedings of the 16th Annual International ACM SIGIR Conference on Research and Development in Information Retrieval},
  pages={160--169},
  year={1993}
}

@inproceedings{qe_sigir_94,
  title={Query expansion using lexical-semantic relations},
  author={Ellen M Voorhees},
  booktitle={Proceedings of the 17th Annual International ACM SIGIR Conference on Research and Development in Information Retrieval},
  pages={61--69},
  year={1994}
}

@article{Li2023TowardsGTE,
  title   = {Towards General Text Embeddings with Multi-stage Contrastive Learning},
  author  = {Zehan Li and Xin Zhang and Yanzhao Zhang and Dingkun Long and Pengjun Xie and Meishan Zhang},
  journal = {arXiv},
  volume  = {abs/2308.03281},
  year    = {2023},
  doi     = {10.48550/arXiv.2308.03281},
  url     = {https://arxiv.org/abs/2308.03281}
}

@inproceedings{word_embeddings_16,
  title     = {Using Word Embeddings for Automatic Query Expansion},
  author    = {Roy, Dwaipayan and Paul, Debjyoti and Mitra, Mandar and Garain, Utpal},
  booktitle = {Neu-IR'16 SIGIR Workshop on Neural Information Retrieval, July 21, 2016, Pisa, Italy},
  year      = {2016}
}

@inproceedings{word_embeddings_19,
  title     = {Deep Neural Networks for Query Expansion Using Word Embeddings},
  author    = {Imani, Ayyoob and Vakili, Amir and Montazer, Ali and Shakery, Azadeh},
  booktitle = {Advances in Information Retrieval: 41st European Conference on IR Research, ECIR 2019, Cologne, Germany, April 14--18, 2019},
  pages     = {203--210},
  year      = {2019}
}

@inproceedings{zheng-etal-2020-bert,
  title={BERT-QE: Contextualized Query Expansion for Document Re-ranking},
  author={Zheng, Zhi and Hui, Kai and He, Ben and Han, Xianpei and Sun, Le and Yates, Andrew},
  booktitle={Findings of the Association for Computational Linguistics: EMNLP 2020},
  pages={4718--4728},
  year={2020}
}

@article{qe_llm_Rolf,
  title={Query expansion by prompting large language models},
  author={Jagerman, Rolf and Zhuang, Honglei and Qin, Zhen and Wang, Xuanhui and Bendersky, Michael},
  journal={arXiv preprint arXiv:2305.03653},
  year={2023}
}

@inproceedings{qe_llm_Iain,
  title={Generative relevance feedback with large language models},
  author={Mackie, Iain and Chatterjee, Shubham and Dalton, Jeffrey},
  booktitle={Proceedings of the 46th International
ACM SIGIR Conference on Research and Development in Information Retrieval},
  pages={2026–-2031},
  year={2023}
}

@article{qe_llm_query_to_doc,
  title={Query2doc: Query expansion with large language models},
  author={Wang, Liang and Yang, Nan and Wei, Furu},
  journal={arXiv preprint arXiv:2303.07678},
  year={2023}
}

@inproceedings{qe_llm_GaQR,
  title={GaQR: An Efficient Generation-augmented Question Rewriter},
  author={Young, Oliver and Fan, Yixing and Zhang, Ruqing and Guo, Jiafeng and de Rijke, Maarten and Cheng, Xueqi},
  booktitle={Proceedings of the 33rd ACM International Conference on Information and
Knowledge Management},
  pages={4228--4232},
  year={2024}
}

@article{qe_llm_Corpus,
  title={Corpus-Steered Query Expansion with Large Language Models},
  author={Lei, Yibin and Cao, Yu and Zhou, Tianyi and Shen, Tao and Yates, Andrew},
  journal={arXiv preprint arXiv:2402.18031},
  year={2024}
}

@inproceedings{qe_llm_Abe,
  title={LLM-based Query Expansion Fails for Unfamiliar and Ambiguous Queries},
  author={Abe, Kenya and Takeoka, Kunihiro and Kato, Makoto P. and Oyamada, Masafumi},
  booktitle={Proceedings of the 48th International
ACM SIGIR Conference on Research and Development in Information Retrieval},
  pages={3035--3039},
  year={2025}
}

@incollection{rocchio1971relevance,
  title={Relevance Feedback in Information Retrieval},
  author={Rocchio, J. J.},
  booktitle={The SMART Retrieval System: Experiments in Automatic Document Processing},
  editor={Salton, Gerard},
  pages={313--323},
  year={1971},
  publisher={Prentice-Hall}
}

@inproceedings{lavrenko2001relevance,
  title={Relevance-Based Language Models},
  author={Lavrenko, Victor and Croft, W. Bruce},
  booktitle={Proceedings of the 24th Annual International ACM SIGIR Conference on Research and Development in Information Retrieval},
  pages={120--127},
  year={2001}
}

@inproceedings{yee2003faceted,
  title={Faceted Metadata for Image Search and Browsing},
  author={Yee, Ka-Ping and Swearingen, Kirsten and Li, Kevin and Hearst, Marti A.},
  booktitle={Proceedings of the SIGCHI Conference on Human Factors in Computing Systems},
  pages={401--408},
  year={2003}
}

@inproceedings{hearst2006design,
  title={Design Recommendations for Faceted Search Interfaces},
  author={Hearst, Marti A.},
  booktitle={SIGIR Workshop on Faceted Search},
  year={2006}
}

@book{tunkelang2009faceted,
  title={Faceted Search},
  author={Tunkelang, Daniel},
  year={2009},
  publisher={Morgan \& Claypool Publishers}
}

@inproceedings{lewis2020retrieval,
  title={Retrieval-Augmented Generation for Knowledge-Intensive NLP Tasks},
  author={Lewis, Patrick and Perez, Ethan and Piktus, Aleksandra and Petroni, Fabio and Karpukhin, Vladimir and Goyal, Naman and K{"u}ttler, Heinrich and Lewis, Mike and Yih, Wen-tau and Rockt{"a}schel, Tim and Riedel, Sebastian and Kiela, Douwe},
  booktitle={Advances in Neural Information Processing Systems},
  year={2020}
}

@article{kwon2023vllm,
  title={vLLM: Easy, Fast, and Cheap LLM Serving with PagedAttention},
  author={Kwon, Woosuk and Li, Zhuohan and Zhuang, Sheng and Sheng, Ying and Zheng, Lianmin and Yu, Cody and Gonzalez, Joseph E. and Stoica, Ion},
  journal={arXiv preprint arXiv:2309.06180},
  year={2023}
}

@article{hinton2015distilling,
  title={Distilling the Knowledge in a Neural Network},
  author={Hinton, Geoffrey and Vinyals, Oriol and Dean, Jeff},
  journal={arXiv preprint arXiv:1503.02531},
  year={2015}
}

@inproceedings{hsieh2023distilling,
  title={Distilling Step-by-Step: Outperforming Larger Language Models with Less Training Data},
  author={Hsieh, Cheng-Yu and Li, Chun-Liang and Yeh, Chih-Kuan and Nakhost, Hootan and Fujii, Yusuke and Ratner, Alex and Krishna, Ranjay and Ma, Tengyu and Farhadi, Ali and Miller, Tom and others},
  booktitle={Proceedings of the 61st Annual Meeting of the Association for Computational Linguistics (ACL)},
  year={2023}
}

@inproceedings{liu2025powering,
  title={Powering Job Search at Scale: LLM-Enhanced Query Understanding in Job Matching Systems},
  author={Liu, Ping and Shen, Jianqiang and Shen, Qianqi and Yao, Chunnan and Kao, Kevin and Xu, Dan and Arora, Rajat and Zheng, Baofen and Johnson, Caleb and Hong, Liangjie and Wu, Jingwei and Zhang, Wenjing},
  booktitle={Proceedings of the 34th CIKM},
  pages={4971--4975},
  year={2025}
}

@article{bai2022constitutional,
  title={Constitutional {AI}: Harmlessness from {AI} Feedback},
  author={Bai, Yuntao and others},
  journal={arXiv preprint arXiv:2212.08073},
  year={2022},
  doi={10.48550/arXiv.2212.08073}
}

@inproceedings{LinkRetrieval2024,
author = {Shen, Jianqiang and Juan, Yuchin and Liu, Ping and Pu, Wen and Zhang, Shaobo and Shen, Qianqi and Hong, Liangjie and Zhang, Wenjing},
title = {Learning Links for Adaptable and Explainable Retrieval},
year = {2024},
booktitle = {Proceedings of the 33rd CIKM},
pages = {4046–4050},
numpages = {5}
}

@article{son2024llm,
  title={Llm-as-a-judge \& reward model: What they can and cannot do},
  author={Son, Guijin and Ko, Hyunwoo and Lee, Hoyoung and Kim, Yewon and Hong, Seunghyeok},
  journal={arXiv preprint arXiv:2409.11239},
  year={2024}
}

@inproceedings{li2025generation,
  title={From generation to judgment: Opportunities and challenges of llm-as-a-judge},
  author={Li, Dawei and Jiang, Bohan and Huang, Liangjie and Beigi, Alimohammad and Zhao, Chengshuai and Tan, Zhen and Bhattacharjee, Amrita and Jiang, Yuxuan and Chen, Canyu and Wu, Tianhao and others},
  booktitle={Proceedings of the 2025 Conference on Empirical Methods in Natural Language Processing},
  pages={2757--2791},
  year={2025}
}

@inproceedings{juan2025scaling,
  title={Scaling Retrieval for Web-Scale Recommenders: Lessons from Inverted Indexes to Embedding Search},
  author={Juan, Yuchin and Shen, Jianqiang and Zhang, Shaobo and Shen, Qianqi and Johnson, Caleb and Simon, Luke and Hong, Liangjie and Zhang, Wenjing},
  booktitle={Proceedings of the 19th ACM Conference on Recommender Systems},
  pages={1066--1069},
  year={2025}
}

@inproceedings{huang2020embedding,
  title={Embedding-based retrieval in facebook search},
  author={Huang, Jui-Ting and Sharma, Ashish and Sun, Shuying and Xia, Li and Zhang, David and Pronin, Philip and Padmanabhan, Janani and Ottaviano, Giuseppe and Yang, Linjun},
  booktitle={Proceedings of the 26th ACM SIGKDD International Conference on Knowledge Discovery \& Data Mining},
  pages={2553--2561},
  year={2020}
}

@article{oord2018representation,
  title={Representation learning with contrastive predictive coding},
  author={Oord, Aaron van den and Li, Yazhe and Vinyals, Oriol},
  journal={arXiv preprint arXiv:1807.03748},
  year={2018}
}

@inproceedings{cao2007learning,
  title={Learning to rank: from pairwise approach to listwise approach},
  author={Cao, Zhe and Qin, Tao and Liu, Tie-Yan and Tsai, Ming-Feng and Li, Hang},
  booktitle={Proceedings of the 24th ICML},
  pages={129--136},
  year={2007}
}

@article{bai2023qwen,
  title={Qwen technical report},
  author={Bai, Jinze and Bai, Shuai and Chu, Yunfei and Cui, Zeyu and Dang, Kai and Deng, Xiaodong and Fan, Yang and Ge, Wenbin and Han, Yu and Huang, Fei and others},
  journal={arXiv preprint arXiv:2309.16609},
  year={2023}
}

@article{yang2025qwen3,
  title={Qwen3 technical report},
  author={Yang, An and Li, Anfeng and Yang, Baosong and Zhang, Beichen and Hui, Binyuan and Zheng, Bo and Yu, Bowen and Gao, Chang and Huang, Chengen and Lv, Chenxu and others},
  journal={arXiv preprint arXiv:2505.09388},
  year={2025}
}

@article{wei2022chain,
  title={Chain-of-thought prompting elicits reasoning in large language models},
  author={Wei, Jason and Wang, Xuezhi and Schuurmans, Dale and Bosma, Maarten and Xia, Fei and Chi, Ed and Le, Quoc V and Zhou, Denny and others},
  journal={Advances in neural information processing systems},
  volume={35},
  pages={24824--24837},
  year={2022}
}

@inproceedings{kenthapadi2017personalized,
  title={Personalized job recommendation system at linkedin: Practical challenges and lessons learned},
  author={Kenthapadi, Krishnaram and Le, Benjamin and Venkataraman, Ganesh},
  booktitle={Proceedings of the eleventh ACM conference on recommender systems},
  pages={346--347},
  year={2017}
}

@article{feng2025we,
  title={Are We on the Right Way to Assessing LLM-as-a-Judge?},
  author={Feng, Yuanning and Wang, Sinan and Cheng, Zhengxiang and Wan, Yao and Chen, Dongping},
  journal={arXiv preprint arXiv:2512.16041},
  year={2025}
}

\end{document}